\documentclass[12pt]{iopart}
\usepackage{graphicx}
\usepackage{cite}

\def\calH{{\cal H}}
\def\vev#1{\left\langle #1 \right\rangle}
\def\vevapp#1{\left\langle #1 \right\rangle^{\mathrm{MFPA}}}

\begin{document}
\title[Effects of the Tsallis distribution in the linear sigma model]{Effects of the Tsallis distribution in the linear sigma model}
\author{Masamichi Ishihara$^1$}
\address{$^1$Department of Human Life Studies, Koriyama Women's University, Koriyama, 963-8503, Japan}
\ead{m\_isihar@koriyama-kgc.ac.jp}

\begin{abstract}
The effects of the Tsallis distribution which has two parameters, $q$ and $T$,
on physical quantities are studied using the linear sigma model in chiral phase transitions.
The Tsallis distribution approaches the Boltzmann-Gibbs distribution as $q$ approaches $1$.
The parameter $T$ dependences of the condensate and mass for various $q$ are shown, where $T$ is called temperature. 
The critical temperature and energy density are described with digamma function, 
and the $q$ dependences of these quantities and the extension of Stefan-Boltzmann limit of the energy density are shown. 
The following facts are clarified. 
The chiral symmetry restoration at $q>1$ occurs at low temperature,  compared with the restoration at $q=1$. 
The sigma mass and pion mass reflect the restoration. 
The critical temperature decreases monotonically as $q$ increases.  
The small deviation from the Boltzmann-Gibbs distribution results in 
the large deviations of physical quantities, especially the energy density. 
It is displayed from the energetic point of view that the small deviation from the Boltzmann-Gibbs distribution is realized for $q>1$.
The physical quantities are affected by the Tsallis distribution even when $|q-1|$ is small.
\end{abstract}
\maketitle

\section{Introduction}
A power-like distribution appears in various branches of science and has been investigated. 
One of them is called Tsallis distribution, and it has been studied in the last few decades. 
The distribution is one parameter extension of the Boltzmann-Gibbs distribution. 
This distribution has been used to analyze various phenomena \cite{TsallisBook}.
An example is the momentum distribution in high energy collisions.

A problem in high energy collisions is to obtain the momentum distribution of emitted particles.
The momentum distribution is fitted with a Tsallis distribution which has two parameters, $q$ and $T$, 
where $T$ is called temperature in the present paper. 
The Tsallis distribution approaches the Boltzmann-Gibbs distribution as $q$ approaches $1$.
It was shown that the momentum distribution is fitted well by the Tsallis distribution with the parameter $q$ which is close to $1$ 
\cite{Alberico2000,Biyajima2005,Biyajima2006,Wilk07_Tsallis,Wilk2009}. 

There are several origins of the Tsallis distribution.
One of them is the Tsallis entropy\cite{Tsallis1988,Tsallis98_constraints}.  
Other origins of the distribution exist:
temperature fluctuation\cite{Wilk2009,WIlk2000,WIlk2013_prepri}, 
multiplicative noise\cite{WIlk2013_prepri,Biro2005,Hasegawa2006_Physica}, etc\cite{WIlk2013_prepri,Hasegawa2006_Physica}.
Therefore, the Tsallis distribution is a probable distribution. 
The definition of the expectation value of a physical quantity is not always obvious,  
because the origin of the Tsallis distribution is not always the Tsallis entropy, 
as discussed in the study of the Tsallis nonextensive statistics.

The phase transition is an important topic in high energy heavy ion collisions. 
A topic is the study of the equation of state in the nonextensive statistics \cite{Drago2004,Pereira2007}.
The NJL model is often used to study the the phase transition. 
This model was used to study the effects of the nonextensive statistics \cite{Rozynek2009}.
The linear sigma model is also used to study the phase transition at high energies. 
The model is the extension of $x^4$ model, and 
the effects of the Tsallis statistics was studied in the $x^4$ model in quantum mechanics \cite{Ishihara2012,Ishihara2014}. 
The phase transition should be investigated with these models, when the distribution is described with the Tsallis distribution.

The purpose of this paper is to clarify the effects of the Tsallis distribution on physical quantities in chiral phase transitions. 
We assume that the momentum distribution is the Tsallis distribution 
and study the variations of the quantities: the condensate, mass, critical temperature, and energy density.
We employ the linear sigma model and study the $q$ dependences of the quantities. 
The expectation value in the present study is not $q$-expectation value used in the Tsallis nonextensive statistics.
The effects of the distribution on the quantities are investigated under the massless free particle approximation. 
The variations of the quantities are studied when the distribution is the Tsallis distribution,
because the possibility of the Tsallis distribution is presented. 

The expressions of the physical quantities are obtained when the momentum distribution is the Tsallis distribution.
The following things are shown. 
The chiral symmetry restoration at $q > 1$ occurs at low temperature,  compared with the restoration at $q=1$. 
The sigma mass and pion mass also change, reflecting the chiral symmetry restoration.
It is shown that the critical temperature decreases as $q$ increases.
The extension of the Stefan-Boltzmann limit of the energy density is shown. 
The energy density increases remarkably as a function of $q$. 
This fact implies that the deviation of $q$ from $q = 1$ is small for $q>1$.  

The outline of this paper is as follows. 
In section~\ref{sec:expression}, 
the expressions of some quantities are calculated in the linear sigma model,  
when the momentum distribution is the Tsallis distribution.
In section~\ref{sec:numericalresults},  
the numerical values of the quantities are shown for various $q$. 
Section~\ref{sec:discussion} is assigned for discussion and conclusion.

\section{Expressions of the condensate, mass, critical temperature, and energy density}
\label{sec:expression}
Scalar fields $\phi=(\phi_0, \phi_1, \cdots, \phi_{N-1})$ are used and 
the Hamiltonian density of the linear sigma model is given by 
\begin{equation}
\calH = \frac{1}{2} \left( \partial^{0} \phi \right)^2 + \frac{1}{2} \left(\nabla \phi \right)^2 + \frac{\lambda}{4} \left( \phi^2 - v^2 \right)^2 - H \phi_0 , 
\label{eqn:Hamiltonian}
\end{equation}
where 
${\displaystyle (\partial^0 \phi)^2 \equiv \sum_{i=0}^{N-1} ( \partial^0 \phi_i) ^2 }$, 
${\displaystyle (\nabla \phi)^2 \equiv \sum_{i=0}^{N-1} ( \nabla \phi_i) ^2 }$, and 
${\displaystyle \phi^2 \equiv \sum_{i=0}^{N-1} \left( \phi_i \right)^2 }$ . 
The Hamiltonian density is rewritten by inserting the decomposition $\phi_i = \phi_{i\rm{c}} + \phi_{i\rm{h}}$  into equation~\eref{eqn:Hamiltonian},
where $\phi_{i\rm{c}}$ represents the condensate and $\phi_{i\rm{h}}$ is the remaining part. 

The expectation value of the Hamiltonian density is calculated.
The expectation value $\vev{\left( \phi_{i\rm{h}} \right) ^{2s+1}}$  is zero under the free particle approximation, 
where $s$ is a non-negative integer and $\vev{O}$ represents the expectation value of the physical quantity $O$. 
The expectation value of Hamiltonian density under the approximation is given by 
\begin{eqnarray}
\vev{\calH(\phi)} &=& \calH(\phi_{\rm{c}}) + \frac{1}{2} \vev{ \left(\partial^0 \phi_{\rm{h}} \right)^2} + \frac{1}{2} \vev{ \left(\nabla \phi _h\right)^2} 
+  \frac{\lambda}{2} \left(\phi_{\rm{c}}^2 - v^2\right) \vev{\phi_{\rm{h}}^2}  
\nonumber \\  && 
+  \frac{\lambda}{4} \vev{ \left(\phi_{\rm{h}}^2\right)^2}  + \lambda \sum_{j=0}^{N-1} \vev{\phi_{j\rm{h}}^2} \left( \phi_{j\rm{c}} \right)^2 
,
\end{eqnarray}
where ${\displaystyle \phi_{\rm{c}}^2 \equiv \sum_{i=0}^{N-1} \left( \phi_{i\rm{c}} \right) ^2 }$ 
and ${\displaystyle \phi_{\rm{h}}^2 \equiv \sum_{i=0}^{N-1} \left( \phi_{i\rm{h}} \right)^2}$.

The massless free particle approximation (MFPA) \cite{Gavin1994,Ishihara1999} is 
applied to calculate the approximate values of the various physical quantities.
The quantity $\vev{\left( \phi_{j\rm{h}} \right)^2}$ under the approximation is given by 
\begin{equation}
\vevapp{\left( \phi_{j\rm{h}} \right)^2} = \int \ \frac{d\vec{k}}{(2\pi)^2 k} \  \vevapp{a_{j\vec{k}}^{\dag} a_{j\vec{k}} } 
+ \int \ \frac{d\vec{k}}{(2\pi)^2 (2 k) } \qquad (k = |\vec{k}|) , 
\end{equation}
where $\vec{k}$ is the momentum and $a_{j\vec{k}}$ is the annihilation operator for the field $\phi_{j\rm{h}}$. 
The second term is discarded in the present calculation, because this term is the contribution of the vacuum.
The expectation value of $\left( \phi_{j\rm{h}} \right)^2 $ under MFPA is represented as 
\begin{equation}
\vevapp{\left( \phi_{j\rm{h}} \right)^2} = \int \frac{d\vec{k}}{(2\pi)^3} \frac{1}{k} f_{q}(\vec{k}), 
\end{equation}
where $f_{q}(\vec{k})$ is the distribution function.

The Tsallis distribution is useful to describe the momentum distribution 
in the analysis of the experiments of high energy heavy ion collisions. 
Therefore, the Tsallis distribution is assumed as $f_{q}(\vec{k})$,  where $q$ is a parameter introduced in the Tsallis distribution.
The Tsallis distribution for boson in MFPA is given by 
\begin{equation}
f_{q}(\vec{k},T) = \frac{1}{\left[ 1 + (q-1) \left( \frac{k}{T} \right)\right]_{+}^{\frac{1}{(q-1)}} -1}, 
\end{equation}
where $T$ is a parameter, which is just the temperature of Boltzmann-Gibbs distribution when $q$ is equal to $1$.   
Therefore, we call the parameter $T$  temperature in this paper.  
The notation $[x]_{+} $ represents $x$ for $x \ge 0$ and $0$ for $x < 0$. 
We define $K_q(T)$ as  $K_q(T) := \vevapp{\left( \phi_{j\rm{h}} \right)^2}$.
The expectation value of Hamiltonian density $\vevapp{\calH (\phi)}$ is given by 
\begin{eqnarray}
\fl
\vevapp{\calH(\phi)} &=& \calH(\phi_{\rm{c}})  
+ \frac{\lambda}{2} \left\{ \left( N+2 \right) \phi_{\rm{c}}^2 - N v^2 \right\} K_q(T) 
\nonumber \\ && 
+ \frac{1}{2} \vevapp{ \left(\partial^0 \phi_{\rm{h}} \right)^2} 
+ \frac{1}{2} \vevapp{ \left(\nabla \phi_{\rm{h}}\right)^2}  
+ \frac{\lambda}{4} \vevapp{ \left(\phi_{\rm{h}}^2\right)^2} . 
\label{vevapp:H}
\end{eqnarray}
The last three terms in the right-hand side of equation~\eref{vevapp:H} are independent of $\phi_{j\rm{c}}$.

The value of the condensate is given as the solution of the following differential equation 
in the case that the condensate $\phi_{j\rm{c}}$ is uniform and time-independent:
\begin{equation}
\frac{\partial{\vevapp{\calH(\phi)}}}{\partial \phi_{i\rm{c}}} \equiv
\lambda \left[ \phi_{\rm{c}}^2  + \left(N+2\right) K_q (T) - v^2 \right] \phi_{i\rm{c}} - H \delta_{i0} = 0 . 
\label{eqn:extremum}
\end{equation}
The mass squared $m_i^2$ is defined by 
\begin{equation}
m_i^2 := \frac{\partial^2{\vevapp{\calH(\phi)}}}{\partial \phi_{i\rm{c}}^2} 
= \lambda \left[ \phi_{\rm{c}}^2  + 2 \left( \phi_{i\rm{c}} \right)^2 + \left(N+2\right) K_q(T) - v^2 \right]  .
\end{equation}

The potential in equation~\eref{eqn:Hamiltonian} is tilted to the $\phi_0$ direction when $H \neq 0$. 
Therefore, the value of the condensate $\phi_{j\rm{c}}$ for $j \neq 0$ is zero for $H \neq 0$.  
Equation~\eref{eqn:extremum} for $\phi_{0\rm{c}}$ is reduced to the following equation:
\begin{equation}
\left( \phi_{0\rm{c}} \right)^3  + \left[ \left(N+2\right) K_q(T) - v^2 \right] \phi_{0\rm{c}} - H/\lambda = 0.
\end{equation}
The mass squared $m_i^2$ is given by 
\begin{equation}
m_i^2 =  \lambda \left[ \left( 1 + 2 \delta_{i0} \right) \left( \phi_{0\rm{c}} \right) ^2  + \left(N+2\right) K_q(T) - v^2 \right]  . 
\end{equation}

Once $K_q(T)$ is given, the condensate $\phi_{0\rm{c}}$ and mass $m_i$ are obtained. 
The quantity $K_q(T)$ is expressed with digamma function $\psi(x)$:
\begin{equation}
K_q(T)  = \frac{I(1,-1)}{2\pi^2}  = \frac{T^2}{2\pi^2 (q-1)} \bigg\{ \psi(2-q) - \psi(3-2q)\bigg\}   \quad (q < 3/2) , 
\end{equation}
where $I(1,-1)$ is given by equation~\eref{eqn:I,1,-1} in \ref{sec:integrals}.
The restriction of $q$ comes from the condition that the integral converges.
The quantity $K_q(T)$ approaches $T^2/12$ as $q$ approaches $1$, as is expected \cite{Gavin1994,Ishihara1999}.

As is well-known,  the critical temperature $T_{\rm{c}}$ cannot be defined definitely when $H \neq 0$. 
In the present study, the critical temperature is the temperature at which a local minimum of the potential vanishes:
a local minimum and  local maximum  merge.
This critical temperature $T_{\rm{c}}(q)$ is given by 
\begin{eqnarray}
\fl
T_{\rm{c}}(q) &=& \sqrt{\frac{2\pi^2}{(N+2)} \left( \frac{q-1}{\psi(2-q) - \psi(3-2q)} \right) 
                            \left[ v^2 - \frac{3}{4} \left(\frac{4H}{\lambda}\right)^{2/3} \right]}
\qquad ( q < 3/2 ) . 
\end{eqnarray}
Therefore, the ratio $T_{\rm{c}}(q)/T_{\rm{c}}(q=1)$ is 
\begin{equation}
\frac{T_{\rm{c}}(q)}{T_{\rm{c}}(q=1)} = \sqrt{\frac{\pi^2}{6} \left( \frac{q-1}{\psi(2-q) - \psi(3-2q)} \right)}  \qquad\qquad ( q < 3/2 ) .  
\end{equation}
This ratio is independent of the number of fields $N$. 

The mass of the field $\phi_{0\rm{c}}$ has a minimum as a function of the temperature. 
The temperature at the minimum, $T_{0}^{*}(q)$,   is calculated by differentiating $\left(m_{0}\right)^2$. 
The ratio $T_0^{*}(q)/ T_{\rm{c}}(q)$ is 
\begin{equation} 
\frac{T_0^{*}(q)}{T_{\rm{c}}(q)} = \sqrt{ \frac{ v^2 + 3 \left( \frac{H}{4\lambda} \right)^{2/3}}{ v^2 - \frac{3}{4} \left( \frac{4H}{\lambda} \right)^{2/3}}} .  
\end{equation} 
This ratio is independent of the parameter $q$ and the number of fields $N$.
This value is determined only by the parameters of the linear sigma model. 
The mass $m_0$ and condensate $\phi_{0\rm{c}}$ at $T_0^{*}$ are given by 
\numparts
\begin{eqnarray}
&& m_0(T_0^{*})  = \sqrt{6\lambda} \left( \frac{H}{4 \lambda}\right)^{1/3}  , 
\label{eqn:m0:min} \\
&& \phi_{0\rm{c}}(T_0^{*}) = \left( \frac{H}{4 \lambda}\right)^{1/3} . 
\end{eqnarray}
\endnumparts
The mass of the field $\phi_{j\rm{c}}$ $(j \neq 0)$  does not have extremum at $T \neq 0$.

The energy density $\varepsilon$ of free particles for one field and pressure density $p$ are given by
\numparts
\begin{eqnarray}
&&\varepsilon = \int \frac{d\vec{k}}{(2\pi)^3} \omega(\vec{k},T) f_q(\vec{k},T) , \\
&&p = \int \frac{d\vec{k}}{(2\pi)^3} \frac{k^2}{3 \omega(\vec{k},T)} f_q(\vec{k},T) , 
\end{eqnarray}
\endnumparts
where $\omega(\vec{k},T)$ is the energy of a particle.
The relation $\varepsilon = 3p$ is also hold in the massless case. 
The energy density divided by $T^4$ for $N$ fields is calculated with equation~\eref{eqn:I,3,-1}  in \ref{sec:integrals}:
\begin{eqnarray}
\frac{\varepsilon(q)}{T^4} &=& \frac{N}{2\pi^2 (q-1)^3}  \bigg\{ \Big[ \psi(2-q) - \psi(5-4q) \Big] 
\nonumber \\ && \qquad\qquad 
- 3 \Big[ \psi(3-2q) - \psi(4-3q) \Big] \bigg\} 
\qquad\qquad (q < 5/4) . 
\end{eqnarray}
The right-hand side is independent of the temperature. 
This is the extension of the Stefan-Boltzmann limit of the energy density. 
The restriction of $q$ also comes from the condition that the integral converges.
The ratio $\varepsilon(q)/\varepsilon(q=1)$ is given by 
\begin{eqnarray}
\frac{\varepsilon(q)}{\varepsilon(q=1)} 
&= \frac{15}{2\pi^4 (q-1)^3}  \bigg\{ \Big[ \psi(2-q) - \psi(5-4q) \Big] 
\nonumber \\ & \qquad\qquad 
- 3 \Big[ \psi(3-2q) - \psi(4-3q) \Big] \bigg\} \qquad\qquad (q < 5/4) . 
\end{eqnarray}

The numerical values of the above quantities are shown in the next section.

\section{Numerical results}
\label{sec:numericalresults}
In this section, the values of the physical quantities are calculated numerically
under the  approximation, MFPA,  discussed in the previous section. 
The number of the fields $N$ is set to 4.
The fields, $\phi_{0}$ and $\phi_{j}$ $(j = 1,2,3)$, are interpreted as the sigma field and  pion fields, respectively. 
The values of the parameters of the linear sigma model are set to $\lambda=20$, $v=87.4\mathrm{MeV}$, and $H=(119 \mathrm{MeV})^3$. 
At $T=0$, the parameters generate  $m_0=600\mathrm{MeV}$, $m_j=135\mathrm{MeV} (j =1, 2, 3)$, and 
pion decay constant $f_{\pi}=92.5\mathrm{MeV}$.

\begin{figure*}
\begin{center}
\includegraphics[width=0.32\textwidth]{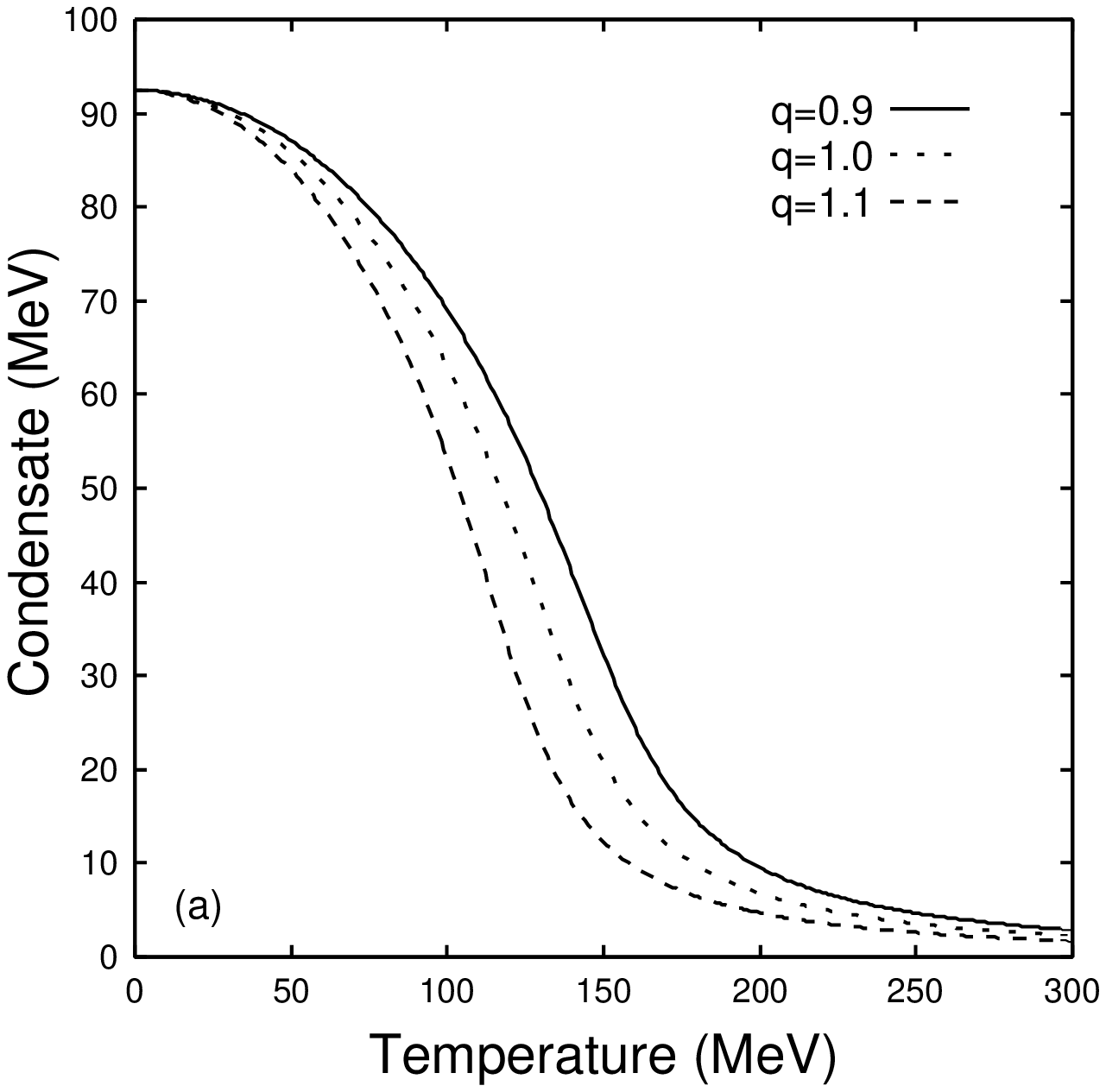}
\includegraphics[width=0.32\textwidth]{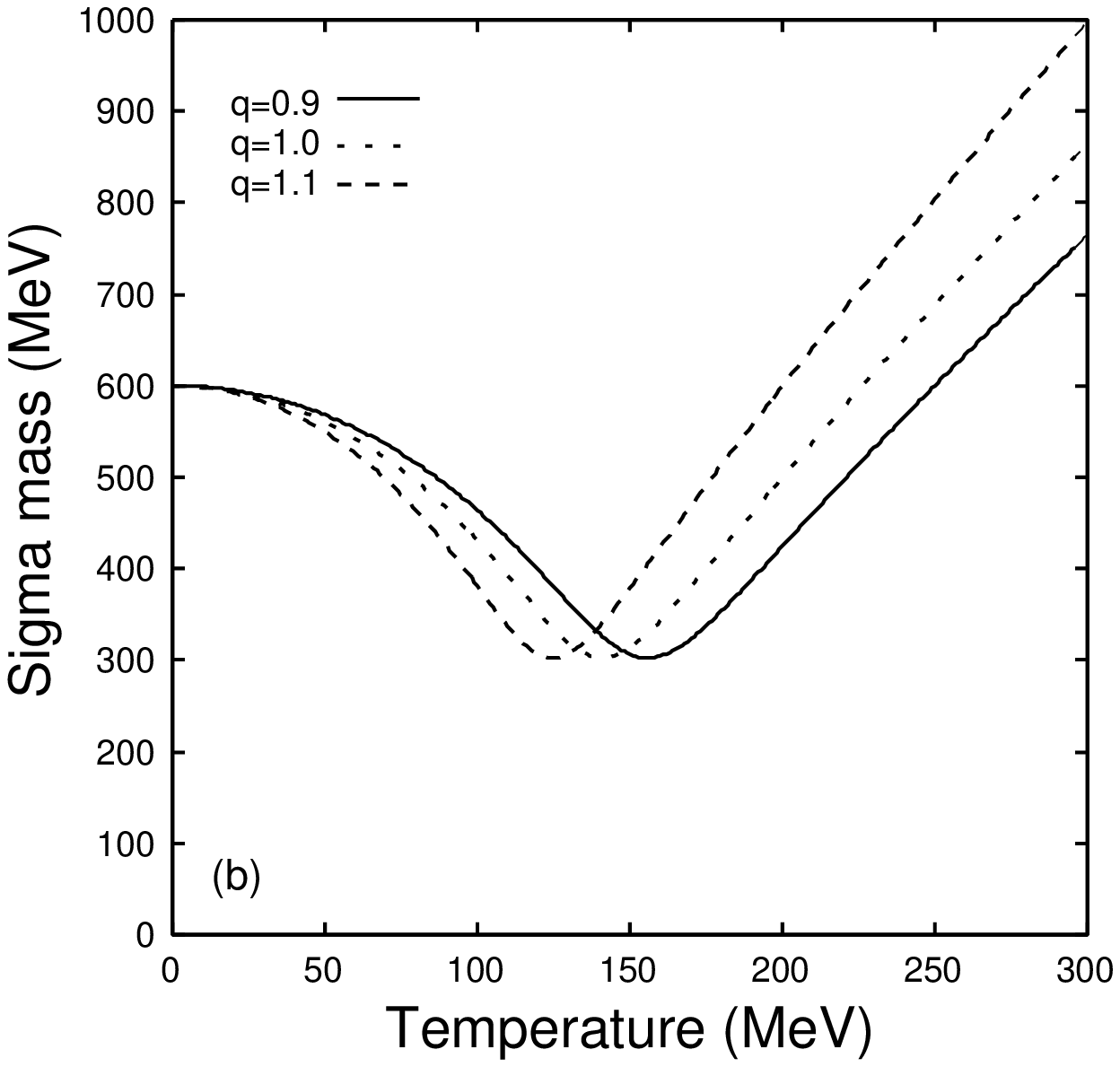} 
\includegraphics[width=0.32\textwidth]{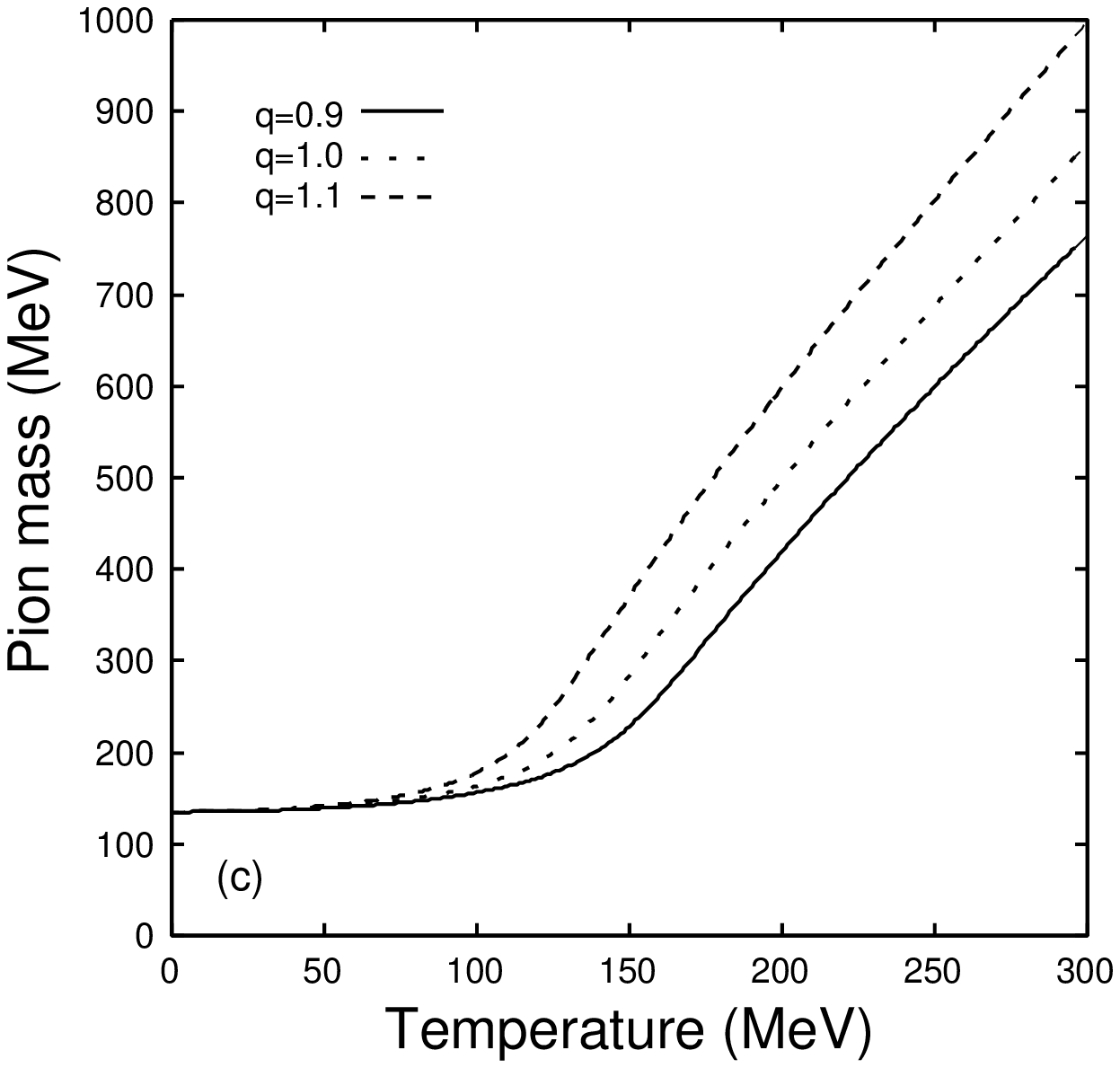}
\end{center}
\caption{Temperature dependences of (a) the condensate, (b) sigma mass, and (c) pion mass for $q=0.9$, $1.0$, and $1.1$. }
\label{fig:vev:mass}
\end{figure*}

Firstly, the temperature dependences of the condensate $\phi_{0\rm{c}}$, sigma mass  $m_0$, and pion mass $m_j$ $(j = 1,2,3)$, 
are shown for $q=0.9$, $1.0$, and $1.1$.
These temperature dependences at $q=1$ are well-known. 
Figure~\ref{fig:vev:mass}(a) shows the temperature dependences of the condensate for $q=0.9$, $1.0$, and $1.1$. 
It is shown that the temperature dependence at $q \neq 1$ is similar to that at $q=1$. 
High temperature is required to restore the symmetry for small $q$.
This $q$ dependence is easily explained by the fact that 
the ratio of the number of the particles with large momentum at $q>1$ is larger than that at $q=1$.
The difference between different $q$ in the condensate can be seen at $T \sim T_{\rm{c}}$. 
Figure~\ref{fig:vev:mass}(b) shows the temperature dependence of the sigma mass  for $q=0.9$, $1.0$, and $1.1$. 
The sigma mass has a minimum and the mass at the minimum is independent of $q$.
The mass at the minimum is 302.5 MeV from equation~\eref{eqn:m0:min} for the present values of the parameters, $\lambda$, $v$, and $H$.
This value is also obtained from the numerical calculation of $m_0$ which is given in figure~\ref{fig:vev:mass}(b).
Figure~\ref{fig:vev:mass}(c) shows the temperature dependence of the pion mass  for $q=0.9$, $1.0$, and $1.1$. 
Contrarily, the pion mass does not have the extremum at $T \neq 0$.  
The difference of the mass between different $q$ can be seen at high temperature at which the symmetry is (partially) restored.

\begin{figure}
\begin{center}
\includegraphics[width=0.4\textwidth]{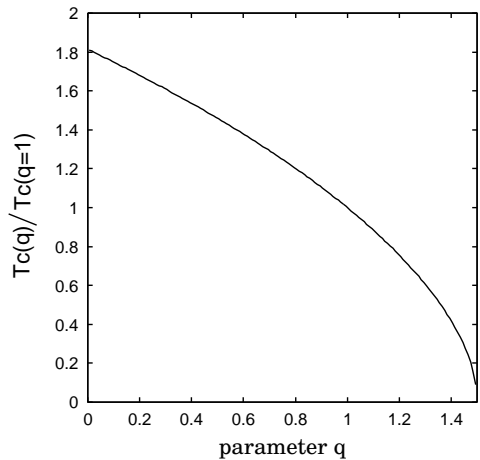}
\end{center}
\caption{The ratio $T_{\rm{c}}(q)/T_{\rm{c}}(q=1)$ as a function of $q$, where $T_{\rm{c}}(q)$ is the critical temperature.}
\label{ratio:TcTcq=1}
\end{figure}

Secondly, the behavior of  the critical temperature $T_{\rm{c}}$ as a function of $q$ is shown. 
The ratio $T_{\rm{c}}(q)/T_{\rm{c}}(q=1)$ is shown in figure~\ref{ratio:TcTcq=1}. 
The parameter $q$ is less than $3/2$ as discussed in the previous section. 
The critical temperature decreases as $q$ increases in the figure. 
The critical temperature $T_{\rm{c}}(q=1)$  in the present case is approximately $90\mathrm{MeV}$ \cite{Ishihara1999_2} . 
Therefore, the critical temperature at $q=1.1$, $T_{\rm{c}}(q=1.1)$, is approximately $10\mathrm{MeV}$ lower than $T_{\rm{c}}(q=1)$.
The ratio of the temperature $T_0^{*}(q)$ to the critical temperature $T_{\rm{c}}(q)$ is approximately $1.57$ in the present case.  
The $q$ dependence of $T_0^{*}(q)$ is the same as that of $T_{\rm{c}}(q)$.
Therefore, the behavior of the  ratio $T_0^{*}(q)/T_{\rm{c}}(q=1)$ is the same, like $T_{\rm{c}}(q)/T_{\rm{c}}(q=1)$.

\begin{figure*}
\begin{center}
\includegraphics[width=0.33\textwidth]{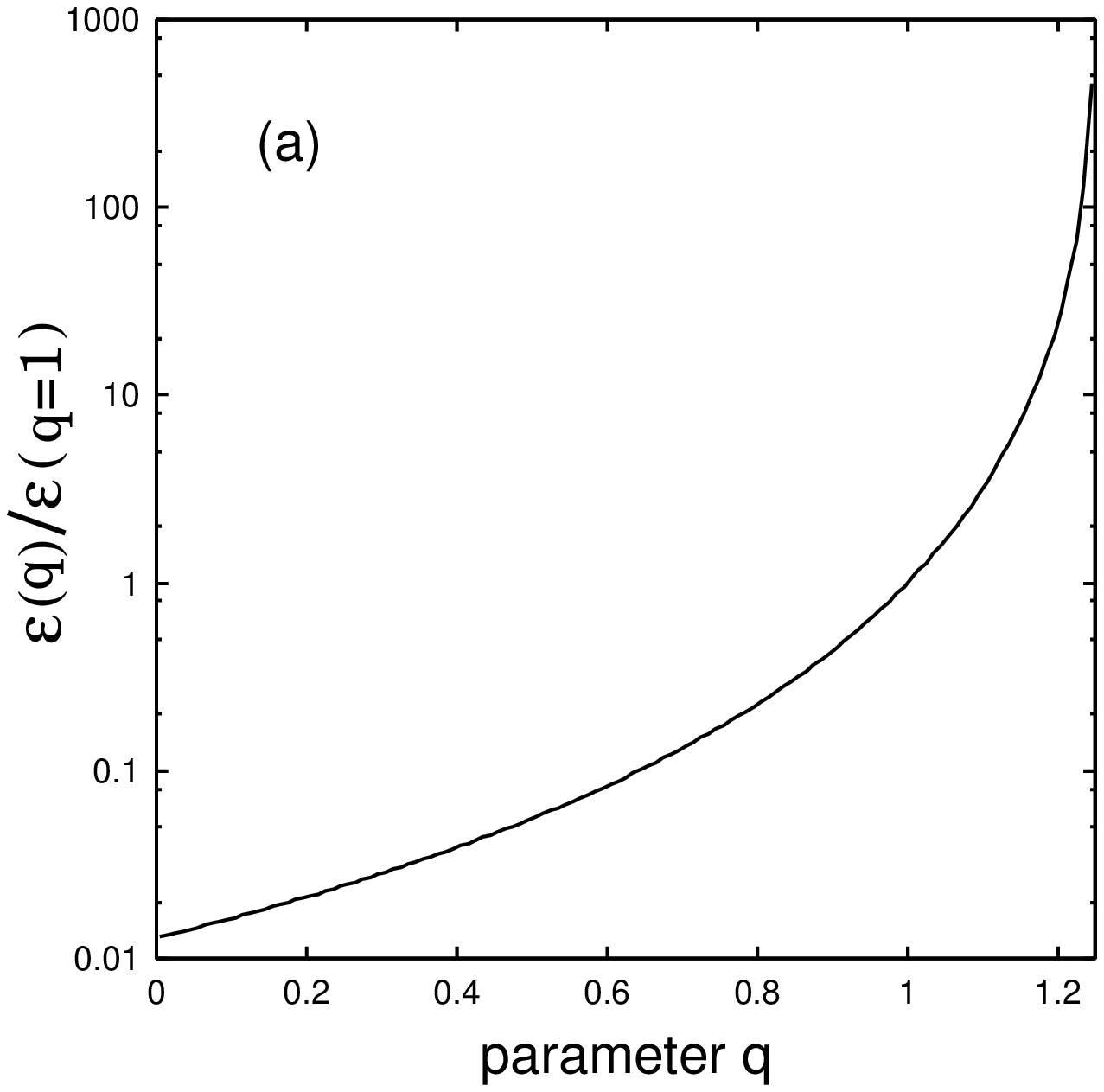}
\includegraphics[width=0.33\textwidth]{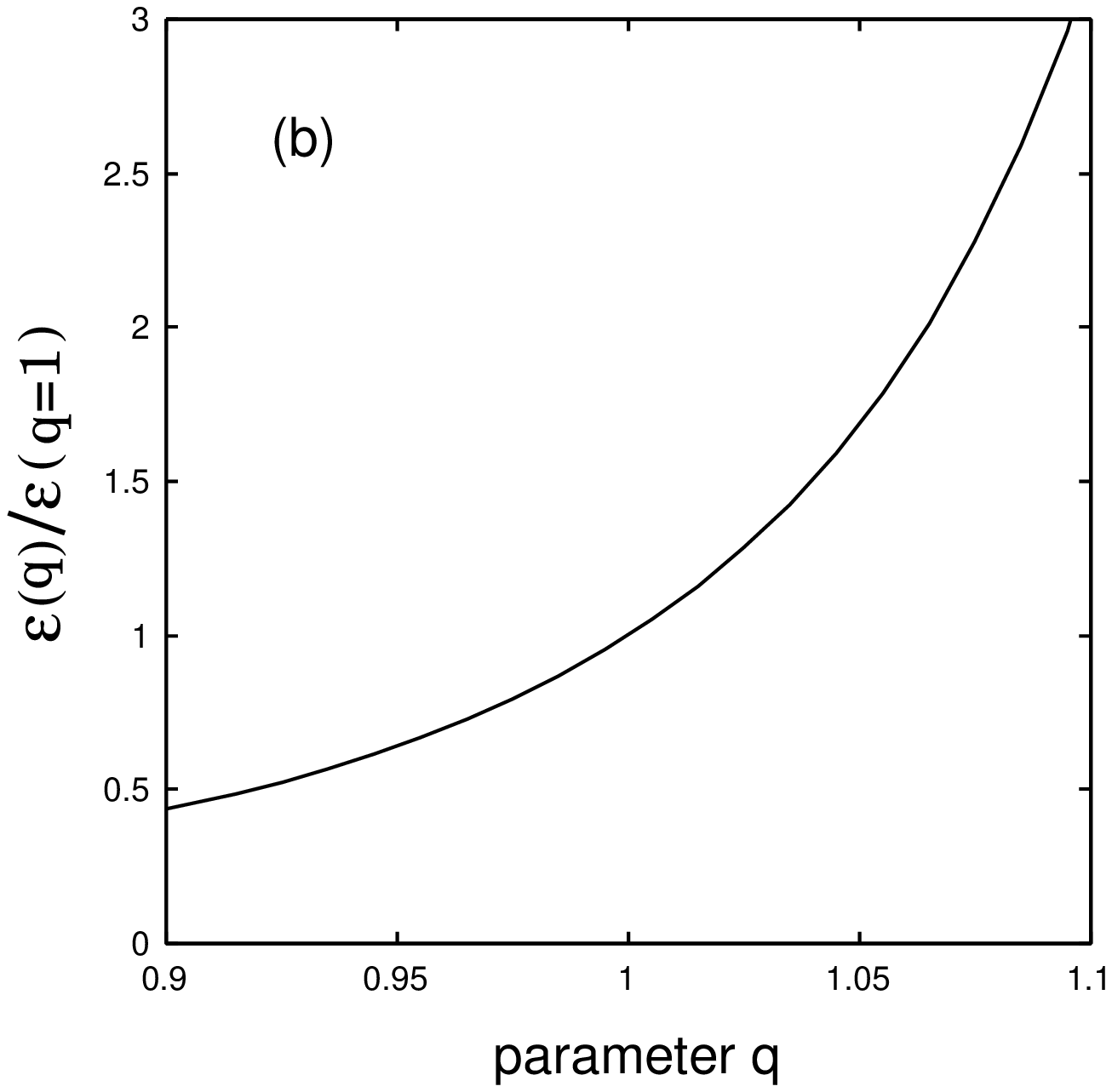}
\end{center}
\caption{Ratio of the energy density $\varepsilon(q)$ to $\varepsilon(q=1)$ in the region of (a) $0 < q < 5/4$ and (b) $0.9 < q < 1.1$.}
\label{ratio:epsq/epsq=1}
\end{figure*}

Finally, 
the ratio $\varepsilon(q)/\varepsilon(q=1)$ is shown in figure~\ref{ratio:epsq/epsq=1}. 
The energy density increases markedly as $q$ increases, as shown in figure~\ref{ratio:epsq/epsq=1}(a).  
The increase of the energy density around $q=1$ is shown in figure~\ref{ratio:epsq/epsq=1}(b).  
The energy density at $q=1.1$, $\varepsilon(q=1.1)$,  is approximately three times larger than $\varepsilon(q=1)$. 
Much energy is stored for $q>1$ even when the Tsallis distribution is close to the Boltzmann-Gibbs distribution.
As shown in figure~\ref{ratio:epsq/epsq=1}(a),  the energy density diverges at $q=5/4$. 
Therefore, the temperature $T(q)$ is restricted, because $q$ is restricted from energetic point of view.
We note the values of the ratio $T_{\rm{c}}(q)/T_{\rm{c}}(q=1)$. 
The ratio $T_{\rm{c}}(q)/T_{\rm{c}}(q=1)$ for $q=3/2$,  $5/4$, and $0$, are $0$, $0.685$, and $1.814$, respectively. 
The restriction caused by the energy density, $q<5/4$, indicates that 
the ratio $T_{\rm{c}}(q)/T_{\rm{c}}(q=1)$ is larger than $T(q=5/4)/T(q=1)$ in figure~\ref{ratio:TcTcq=1}.

\section{Discussion and Conclusion}
\label{sec:discussion}
We studied the effects of the Tsallis distribution in the linear sigma model. 
The Tsallis distribution was assumed as the distribution function, 
because some origins of the Tsallis distribution are suggested and 
the Tsallis distribution describes many phenomena well in various branches of science, such as momentum distribution.
The Tsallis distribution has two parameters, $q$ and $T$, where we call the parameter $T$ temperature. 
The distribution approaches the Boltzmann-Gibbs distribution as $q$ approaches $1$.
The effects of the distribution on physical quantities, such as condensate, were investigated. 
We note that the expectation value in the present study is not the $q$-expectation value used in the Tsallis nonextensive statistics
and that studying the realization of the Tsallis statistics is out of the scope.

At the same temperature, the condensate at $q>1$ is smaller than the condensate at $q=1$ 
and the condensate at $q<1$ is larger than the condensate at $q=1$. 
The behavior depends on the expectation value of the square of a field, $(\phi_{i\rm{h}})^2$, which increases as $q$ increases.
This increase comes from the fact that the tail of the distribution is long for $q>1$.
The parameter $q$ is restricted in the region of $q < 3/2$ in the calculation of the expectation value of $(\phi_{i\rm{h}})^2$.

The increase of the expectation value of $(\phi_{i\rm{h}})^2$  affects the mass.   
The pion mass $m_j$ $(j = 1, 2, 3)$ at $q>1$ is heavier than the mass at $q=1$ 
and the pion mass at $q<1$ is lighter than the mass at $q=1$. 
At high temperature, the sigma mass at $q>1$ is heavier than the mass at $q=1$. 
However, at low temperature, the sigma mass at $q>1$ is lighter than the mass at $q=1$. 
This behavior at low temperature comes from the behavior of the condensate. 
The quantity $\vev{(\phi_{i\rm{h}})^2}$ at $q>1$ is large, compared with the value at $q=1$, 
and the value of the condensate is small at $q>1$ at low temperature, compared with the value at $q=1$. 
The sigma mass at $q>1$ decreases rapidly at low temperature, reflecting the value of the condensate.

The critical temperature $T_{\rm{c}}(q)$ decreases as $q$ increases. 
This behavior depends on the expectation value of the square of a field, $(\phi_{i\rm{h}})^2$, as the condensate does. 
The critical temperature changes even when $q$ changes slightly.
The sigma mass $m_0$ has a minimum at the temperature $T_0^{*}(q)$, 
and the $q$ dependence of the temperature $T_0^{*}(q)$ is the same as that of $T_{\rm{c}}(q)$.
The ratio $T_0^{*}(q) / T_{\rm{c}}(q)$ depends only on the parameters of the linear sigma model. 
The mass $m_0$ at $T_0^{*}(q)$ depends also only on the parameters of the linear sigma model. 

The energy density $\varepsilon(q)$ is proportional to $T^4$.
The ratio $\varepsilon(q)/(N T^4)$ depends only on the parameter $q$ under the massless free particle approximation,
where $N$ is the number of fields.  
The extension of the Stefan-Boltzmann limit for the energy density, $\varepsilon(q)/T^4$, is shown. 
The parameter $q$ is restricted in the region of $q<5/4$ in the calculation of $\varepsilon(q)$.
Physically, the deviation of $q$ from $q=1$ for $q>1$ should be small,  
because the energy density increases remarkably as $q$ increases. 
The critical temperature $T(q)$  is also restricted from the restriction of $q$:
the ratio $T(q)/T(q=1)$  is larger than $T(q=5/4)/T(q=1)$.

The variations of the physical quantities are obviously affected by the Tsallis distribution in spite of small $|q-1|$.  
The remarkable increase of the energy density as a function of $q$ implies that the quantity $|q-1|$ should be small for $q>1$, 
because it is difficult to increase the energy density extremely. 
It is not equal $q$ being $1$ that $|q-1|$ being small.

In summary, 
the expressions of physical quantities, such as critical temperature, are obtained 
when the momentum distribution is the Tsallis distribution which has two parameters, $q$ and $T$, 
where the parameter $T$ is called the temperature in this paper. 
The chiral symmetry restoration at $q>1$ occurs at lower temperature,  compared with the symmetry restoration at $q=1$, 
and the changes of the sigma mass and pion mass reflect the chiral symmetry restoration.
The effects of the Tsallis distribution on the condensate are shown at $T \sim T_{\rm{c}}$,
and the effects of the Tsallis distribution on the mass are shown at $T > T_{\rm{c}}$.
The critical temperature decreases monotonically as $q$ increases.  
The small deviation of $q$ from $q=1$ results in 
the large deviations of the quantities, especially the energy density. 
The extension of the Stefan-Boltzmann limit  of the energy density is shown, and 
it is expected from energetic point of view that the small deviation of $q$ from $q=1$ is realized for $q>1$.
It is not equal $q$ being $1$ that $|q-1|$ being small, and 
the physical quantities are affected by the Tsallis distribution even when $|q-1|$ is small.

The results will be useful to study the phenomena in high energy collisions, 
and the further works related to the Tsallis distribution will be performed in the near future.

\medskip \noindent {\bf References} 

\appendix
\section{Integrals in the calculations}
\label{sec:integrals}
The following integral appears in the calculations:  
\begin{equation}
I(\mu; \xi) := \int_{0}^{\infty} \ dk \ \left(  \frac{k^{\mu}}{ \left[ 1 + (q-1) \beta k \right]_{+}^{1/(q-1)} + \xi} \right) 
\quad (\xi = -1,\ 0,\ 1) , 
\end{equation}
where the parameter $\mu$ is a non-negative integer, and the notation $[x]$ is defined as follows:
\begin{equation}
[x]_{+} = 
\cases{
x & (x $\ge$ 0)\\
0 & (x $<$ 0)
}
.
\end{equation}
The parameter $\xi$ is $-1$ for boson, $0$ for classical particle, and $1$ for fermion.

This integral is represented by changing of variables:
\begin{equation}
I(\mu; \xi) = \frac{1}{\beta^{\mu+1} (q-1)^{\mu}} \int_{0}^{1} \ dy \ \left( \frac{y^{1-q} \left[ y^{1-q} - 1 \right]^{\mu} }{1 + \xi y}  \right) .
\end{equation}
The integral can be represented with digamma  function $\psi(x)$ \cite{Abramowitz,Iwanami_III} which is given by 
\begin{equation}
\psi(x) = \frac{1}{\Gamma(x)} \left( \frac{d\Gamma(x)}{dx} \right). 
\end{equation}
The integrals $I(1,\xi)$ for $q < 3/2$ are calculated:
\newcounter{tmpequation}
\setcounter{tmpequation}{\value{equation}}
\addtocounter{tmpequation}{1}
\renewcommand{\theequation}{A.\arabic{tmpequation}\alph{equation}}
\setcounter{equation}{0}
\begin{eqnarray}
&&I(1,-1) = \frac{1}{\beta^2 (q-1)} \bigg\{ \psi(2-q) - \psi(3-2q)\bigg\} , \label{eqn:I,1,-1}\\
&&I(1,0) = \frac{1}{\beta^2 (q-1)}\left\{ \frac{1}{(3-2q)} - \frac{1}{(2-q)} \right\}, \\
&&I(1,1) = \frac{1}{\beta^2 (q-1)} \left\{ 
\frac{1}{2}  \left[ \psi(2-q) - \psi\left(\frac{3-2q}{2}\right) \right] \right. 
\nonumber \\ && \qquad\qquad\qquad\qquad
\left.  - \frac{1}{2}  \left[ \psi \left(\frac{3-q}{2}\right) - \psi\left( \frac{2-q}{2} \right) \right] 
\right\}  .  
\end{eqnarray}
The integrals $I(3,\xi)$ for $q < 5/4$ are calculated:
\addtocounter{tmpequation}{1}
\setcounter{equation}{0}
\begin{eqnarray}
&& I(3,-1) = \frac{1}{\beta^4 (q-1)^3} \bigg\{ \Big[ \psi(2-q) - \psi(5-4q) \Big] 
  \nonumber \\ && \qquad\qquad\qquad\qquad
  - 3 \Big[ \psi(3-2q) - \psi(4-3q) \Big] \bigg\} , \label{eqn:I,3,-1}\\
&& I(3,0) = \frac{1}{\beta^4 (q-1)^3} \Bigg\{ \left[ \frac{1}{(5-4q)} - \frac{1}{(2-q)}  \right] 
  \nonumber \\ && \qquad\qquad\qquad\qquad
  - 3 \left[ \frac{1}{(4-3q)} - \frac{1}{(3-2q)} \right] \Bigg\} , \\
&& I(3,1) = \frac{1}{\beta^4 (q-1)^3} \bigg\{  \Big[ J_3 - J_0 \Big] -  3 \Big[ J_2 - J_1 \Big] \bigg\} , 
\end{eqnarray}
where $J_p$ is given by 
\renewcommand{\theequation}{A.\arabic{equation}}
\setcounter{equation}{\value{tmpequation}}
\begin{eqnarray}
\fl
J_p = \frac{1}{2} \Bigg\{ \psi\left(\frac{(3+p)-(1+p) q}{2} \right) - \psi\left(\frac{(2+p)-(1+p) q}{2} \right) \Bigg\}  
\qquad
\left( q < \frac{2+p}{1+p}  \right) .
\end{eqnarray}
Some quantities are represented with these expressions.

\end{document}